\documentclass[twocolumn]{aastex63}

\submitjournal{ApJ}

\shorttitle{The inverted ejecta layers in Cassiopeia~A}
\shortauthors{Tsuchioka et al.}
\graphicspath{{./}{figures/}}

\usepackage{longtable}
\usepackage{multirow}
\usepackage{threeparttable}
\usepackage{color}
\usepackage{lineno}
\usepackage{xcolor}

\usepackage{graphicx}

\begin{document}

\title{X-ray Studies of the Inverted Ejecta Layers in the Southeast Area of Cassiopeia~A}

\correspondingauthor{Tomoya Tsuchioka}
\email{t.tsuchioka@rikkyo.ac.jp}

\author[0000-0002-8604-1641]{Tomoya Tsuchioka}
\affil{Department of Physics, Rikkyo University, 3-34-1 Nishi Ikebukuro, Toshima-ku, Tokyo 171-8501, Japan}

\author[0000-0001-9267-1693]{Toshiki Sato}
\affil{Department of Physics, Rikkyo University, 3-34-1 Nishi Ikebukuro, Toshima-ku, Tokyo 171-8501, Japan}

\author[0000-0003-4808-893X]{Shinya Yamada}
\affil{Department of Physics, Rikkyo University, 3-34-1 Nishi Ikebukuro, Toshima-ku, Tokyo 171-8501, Japan}

\author{Yasunobu Uchiyama}
\affil{Department of Physics, Rikkyo University, 3-34-1 Nishi Ikebukuro, Toshima-ku, Tokyo 171-8501, Japan}



\begin{abstract}

The central strong activities in core-collapse supernovae expect to produce the overturning of the Fe- and Si/O-rich ejecta during the supernova explosion based on multi-dimensional simulations. X-ray observations of the supernova remnant Cassiopeia~A have indicated that the Fe-rich ejecta lies outside the Si-rich materials in the southeastern region, which is consistent with the hypothesis on the inversion of the ejecta.
We investigate the kinematic and nucleosynthetic properties of the inverted ejecta layers in detail to understand its formation process using the data taken by the Chandra X-ray Observatory.
Three-dimensional velocities of Fe- and Si/O-rich ejecta are obtained as $>$4,500 km s$^{-1}$ and $\sim$2,000--3,000 km s$^{-1}$, respectively, by combining proper motion and line-of-sight velocities, indicating that the velocity of the Si/O-rich ejecta is slower than that of the Fe-rich ejecta since the early stage of the explosion.
To constrain their burning regime, 
the Cr/Fe mass ratios are evaluated as \(0.51_{-0.10}^{+0.11}\)\% in the outermost Fe-rich region 
and $1.24 ^{+0.19}_{-0.20}$\% in the inner Fe/Si-rich region, suggesting that the complete Si burning layer is invertedly located to the incomplete Si burning layer.
All the results support the ejecta overturning at the early stages of the remnant's evolution or during the supernova explosion of Cassiopeia~A.

\end{abstract}



\section{Introduction} \label{sec:intro}
Stars with masses higher than about $8~$\(M_{\odot}\) undergo a gravitational core-collapse supernova (CC SN) explosion at the final stage of their evolution. Currently, the explosion mechanism of CC SNe is one of the unsolved problems in astrophysics. In theory, central asymmetric effects during the explosion are widely believed to help such massive stars to explode \citep[e.g.,][]{1995ApJ...450..830B,1996A&A...306..167J,1999ApJ...524L.107K,2000ApJ...541.1033F,2003ApJ...584..971B,2009ApJ...691.1360T,2012ApJ...749...98T,2012ARNPS..62..407J,2016ARNPS..66..341J,2021Natur.589...29B}. However, it is difficult to investigate the asymmetric mechanisms only from SN observations.

Observations of supernova remnants (SNRs) provide us a great opportunity to examine those asymmetries during SN explosions. Especially in young SNRs, the distribution of synthesized elements and their kinematics provide us unique information on the asymmetries at the explosion, such as neutrino-star kick, ejecta mixing, and so on \citep[e.g.,][]{2007ApJ...670..635W,2011ApJ...732..114L,2013ApJ...764...50L,2017ApJ...844...84H,2018ApJ...856...18K,2020ApJ...889..144H,2021A&A...646A..82P,2021ApJ...912..131T,2021Natur.592..537S}. 
Here, the galactic supernova remnant Cassiopeia~A (Cas~A) is one of the most well-studied samples to discuss it. The highly asymmetric ejecta distributions of the remnant were observed in detail at multiple wavelengths \citep[e.g.,][]{2001AJ....122.2644F,2006ApJ...636..859F,2009ApJ...693..713S,2010ApJ...725.2038D,2010ApJ...725.2059I,2014Natur.506..339G,2012ApJ...757..126I,2014ApJ...785....7D,2014MNRAS.441.2996A,2014ApJ...789..138P,2015Sci...347..526M,2016ApJ...818...17F,2017ApJ...834...19G,2022arXiv220103753I}, which can be a unique tool to test the asymmetric effects by comparing them with theoretical calculations \citep[e.g.,][]{2008ApJ...686..399S,2008ApJ...677.1091W,2016ApJ...822...22O,2017ApJ...842...13W,2021A&A...645A..66O}.

One of the peculiar structures observed in Cassiopeia~A would be the Fe-rich ejecta located at the tip of the southeastern region \citep{2000ApJ...528L.109H,2000ApJ...537L.119H,2003ApJ...597..362H,2004NewAR..48...61V,2021Natur.592..537S}.
In general, the Fe-rich materials are synthesized at the  innermost region of the SN explosions via explosive Si burning \citep[e.g.,][]{1973ApJS...26..231W,1996ApJ...460..408T}.

On the other hand, the Fe-rich ejecta in the remnant is located outside of the other lighter elements such as Si and O in the 2D projected image, suggesting that some strong central activities pushed this structure outward  \citep[e.g.,][]{2000ApJ...528L.109H,2021Natur.592..537S}. For example, it has been theoretically suggested that explosions driven by bipolar jets could produce such ``overturning" of the ejecta layers \citep[e.g.,][]{2003ApJ...598.1163M}.
Also, recent multi-dimensional SN simulations have indicated that the neutrino-driven mechanism can reproduce well the large clumps of the Fe-rich ejecta observed in the remnant \citep[][]{2017ApJ...842...13W}. Thus, this peculiar feature is possibly related to the central mechanism of the explosion.

While the observational and theoretical studies have supported that the inverted layers are the result of the central strong activities during the explosion, our understanding of the process of the spatial inversion is still being updated. \cite{2010ApJ...725.2038D} have constructed a comprehensive 3D model of the remnant using infrared and X-ray Doppler velocity measurements, which dismissed the overturning during the explosion as the origin of the southeastern Fe-rich structures \citep[see also][]{2013ApJ...772..134M}. The authors have proposed that ``pistons'' of faster than average ejecta can explain well that the Fe-rich ejecta in this region occupies a ``ring'' of the Si-rich ejecta. In this picture, emissions from the piston itself fade gradually after the reverse-shock passage leaving such a ring-like structure, and then the ejecta layers may remain intact or the piston may ``breakthrough'' the outer layers. This could have produced the observed structures where the Fe is located at the tip of the southeastern region even without the inversion of between Fe and Si layers at the explosion.
Recent 3D SNR simulations \citep{2016ApJ...822...22O,2021A&A...645A..66O} have demonstrated that the Fe-rich plumes that were created during the initial stages of the SN explosion can reproduce the inverted ejecta layers in the remnant, where a similar way to the pistons seems to work during the SNR evolution for producing the spatial inversion of the ejecta layers at the age of the remnant.  

In this paper, we aim to investigate both kinematic and nucleosynthetic properties of the inverted ejecta layers to understand its formation process in more detail using X-ray observations of Cas A. From proper motion and Doppler velocity measurements, we can quantitatively assess the current 3D velocity and deceleration of each ejecta layer, which would allow us to discuss the initial kinematics of each component at the explosion. Also, with the constraints of the burning regime of each ejecta layer, we can discuss the inversion process from a different perspective than the kinematic properties. 
The previous observations have not reported the existence of ejecta from the incomplete Si burning layer. This would be a notable point because the incomplete Si burning layer must be located just above the complete Si burning layer in spherically symmetric SN models \citep[e.g.,][]{1995ApJS..101..181W,1996ApJ...460..408T}. On the other hand, we can see only the products from the complete Si burning (i.e., $\alpha$-rich freeze out) regime at the tip of the southeastern Fe-rich structures \citep[e.g.,][]{2003ApJ...597..362H,2021Natur.592..537S}. It would be possible that the strong asymmetric activities during the explosion caused this missing layer to be located behind the complete Si burning layer \citep[e.g.,][]{2003ApJ...598.1163M}, but it needs to be verified with observations. In particular, the inversion of these innermost layers may reflect the overturning at a very early stage of the SN/SNR evolution (at least, earlier than the inversion of the Fe/Si layers). Thus, we also aim to search for the missing incomplete Si burning layer in the southeastern region in this study.

This paper is organized as follows. The next section summarizes the observations used and the data reduction procedures applied to the data. In Section 3, we present our imaging and spectroscopic analysis of the data and results in the inverted layers of Cas~A. The discussion section (Section 4) studies the kinematics and nucleosynthetic properties of the inverted layers and the final section summarizes the article. Throughout this article, uncertainties are quoted at the 90\% confidence level, unless explicitly stated otherwise.

\begin{deluxetable*}{ccccccc}[t!]
\tablecaption{Basic information on the Chandra observations of Cas~A used in the analysis.}
\tablehead{
 \colhead{Obs. ID} & \colhead{Obs. Start} & \colhead{Exposure} & \colhead{Detector}& \colhead{RA}& \colhead{Dec}& \colhead{Roll}\\ 
\colhead{}  &\colhead{(yyyy mm dd)} &\colhead{(ks)} &\colhead{} & \colhead{(deg)} & \colhead{(deg)} & \colhead{(deg)}
}
\startdata
\hline
114&  2000 Jan 30& 49.9 &ACIS-S &  350.9159 & 58.7926 & 323.3801\\\hline
4634& 2004 Apr 28& 148.6& ACIS-S & 350.9047 & 58.8455 & 59.2239 \\
4635&  2004 May 1& 135.0& ACIS-S & 350.9048 & 58.8455 & 59.2237\\
4636&  2004 Apr 20& 143.5& ACIS-S & 350.9129 & 58.8412 & 49.7698 \\
4637 &  2004 Apr 22& 163.5 &ACIS-S& 350.9131 & 58.8414 & 49.7665 \\
4638&  2004 Apr 14 & 164.5 &ACIS-S& 350.9196 & 58.8365 & 40.3327 \\
4639&  2004 Apr 25 & 79.0 &ACIS-S& 350.9132 & 58.8415 & 49.7666 \\
5196&  2004 Feb 8 & 49.5 &ACIS-S& 350.9129 & 58.7933 & 325.5035\\
5319&  2004 Apr 18 & 42.3 &ACIS-S& 325.5035 & 58.8411 & 49.7698 \\
5320&  2004 May 5 & 54.4 &ACIS-S& 350.8988 & 58.8480 & 65.1350 \\\hline
19606&  2019 May 13 & 37.6 &ACIS-S& 350.8854 & 58.8559 & 75.1398 \\
\enddata
\label{tab:chandra_dataset}
\end{deluxetable*}

\section{Observations} \label{sec:obs}

The Advanced CCD Imaging Spectrometer (ACIS) of Chandra has observed Cas~A multiple times since its launch in 1999 \citep[e.g.,][]{2000ApJ...528L.109H,2000ApJ...537L.119H,2014ApJ...789..138P,2018ApJ...853...46S}. The observational data used in this study are summarized in Table~1. In the image analysis, we used ACIS data observed in 2000 (PI: Holt, S) and 2019 (PI: Patnaude, D). For the spectral analysis, we used data from 2004 (PI: Hwang, U) for a total of $\sim $1 Ms \citep[e.g.,][]{2004ApJ...615L.117H}. We used {\tt merge\_obs} to combine the observations of the nine ObsIDs in 2004.

We reprocessed the data using {\tt chandra\_repro} in CIAO 4.13 with CALDB 4.9.5. Before performing the image analysis, we aligned the two sets of data using the coordinates of the Central Compact Object (CCO). It is the only available common point source in the two images. The coordinates of the CCO of each image were obtained with {\tt wavdetect}, and then the position offsets were computed with {\tt wcs\_match}. Using this offset value, we recreated the image to be used for image analysis by updating the coordinates of the event file with {\tt wcs\_update}.

\begin{figure}[t] 
 \centering
 \includegraphics[width=8cm]{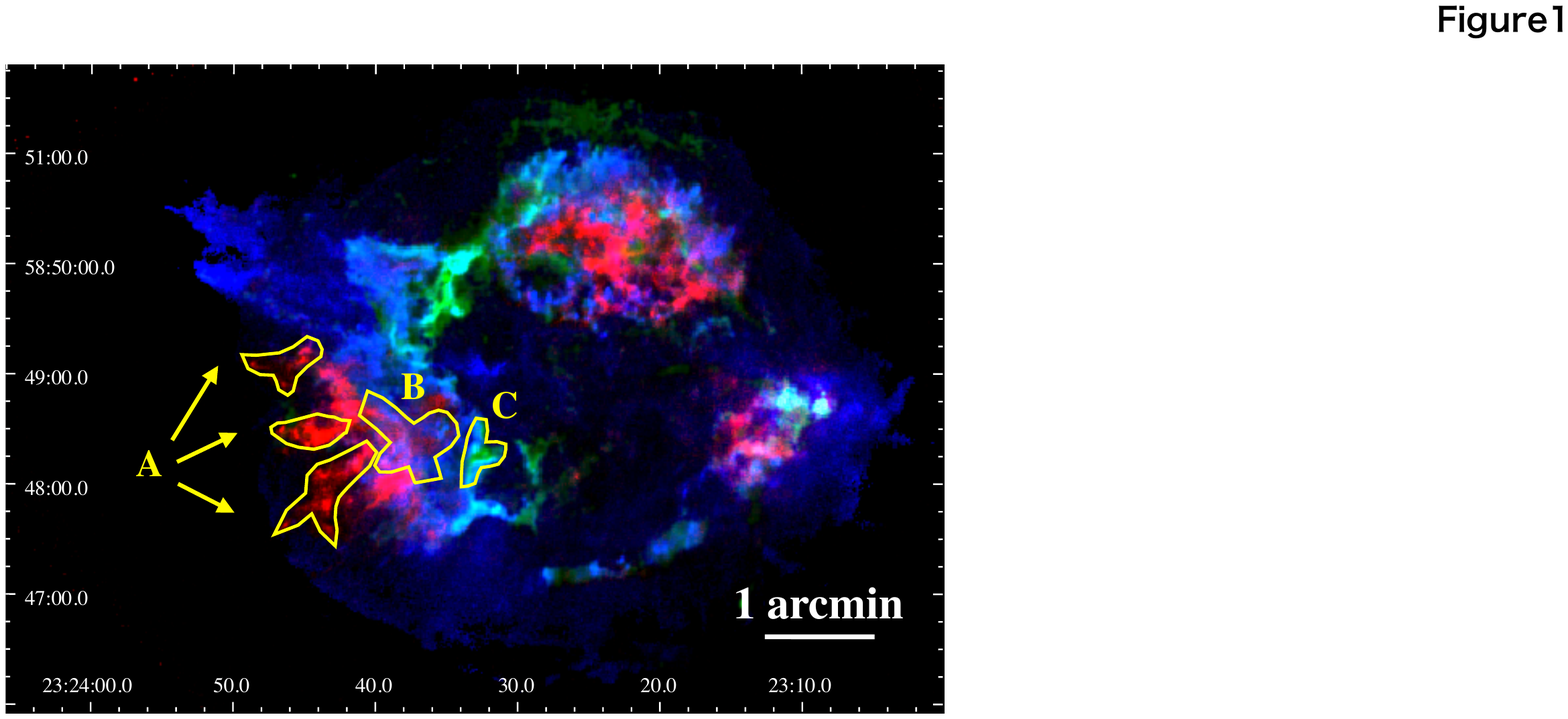}
 \caption{X-ray image of Cas~A taken by Chandra. The red and blue colors highlight Fe-rich (6.3-6.9 keV) and O-rich (0.5-0.7 keV/0.9-1.0 keV) emission regions, respectively. The ratio map of the Si/Mg band (1.8-2.1 keV/1.2-1.6 keV) is shown in green color. Solid yellow contours show the regions used for the spectral analysis.}
\end{figure}

\section{Results} \label{sec:results}
The asymmetric distribution of elements in Cas~A can be seen in the intensity maps of X-ray lines from the shocked ejecta. Figure~1 shows X-ray images of Cas~A, where the Fe-rich, O-rich, and Si-rich ejecta are highlighted in red, green, and blue, respectively. 
In the southeastern region, we can see the inverted ejecta layers reported in the previous studies \citep[e.g.,][]{2000ApJ...528L.109H,2000ApJ...537L.119H,2004NewAR..48...61V}, where we defined the Fe-rich, Fe/Si-rich, and Si/O-rich regions as region A, B, and C, respectively (Figure~1). 
In this section, we attempt two types of analysis: (1) image analysis for determining the kinematic properties (i.e., proper motions) of the ejecta and (2) spectral analysis for measuring the elemental composition to constrain the burning regime of each ejecta layer.

\begin{figure}[t] 
 \centering
 \includegraphics[width=8cm]{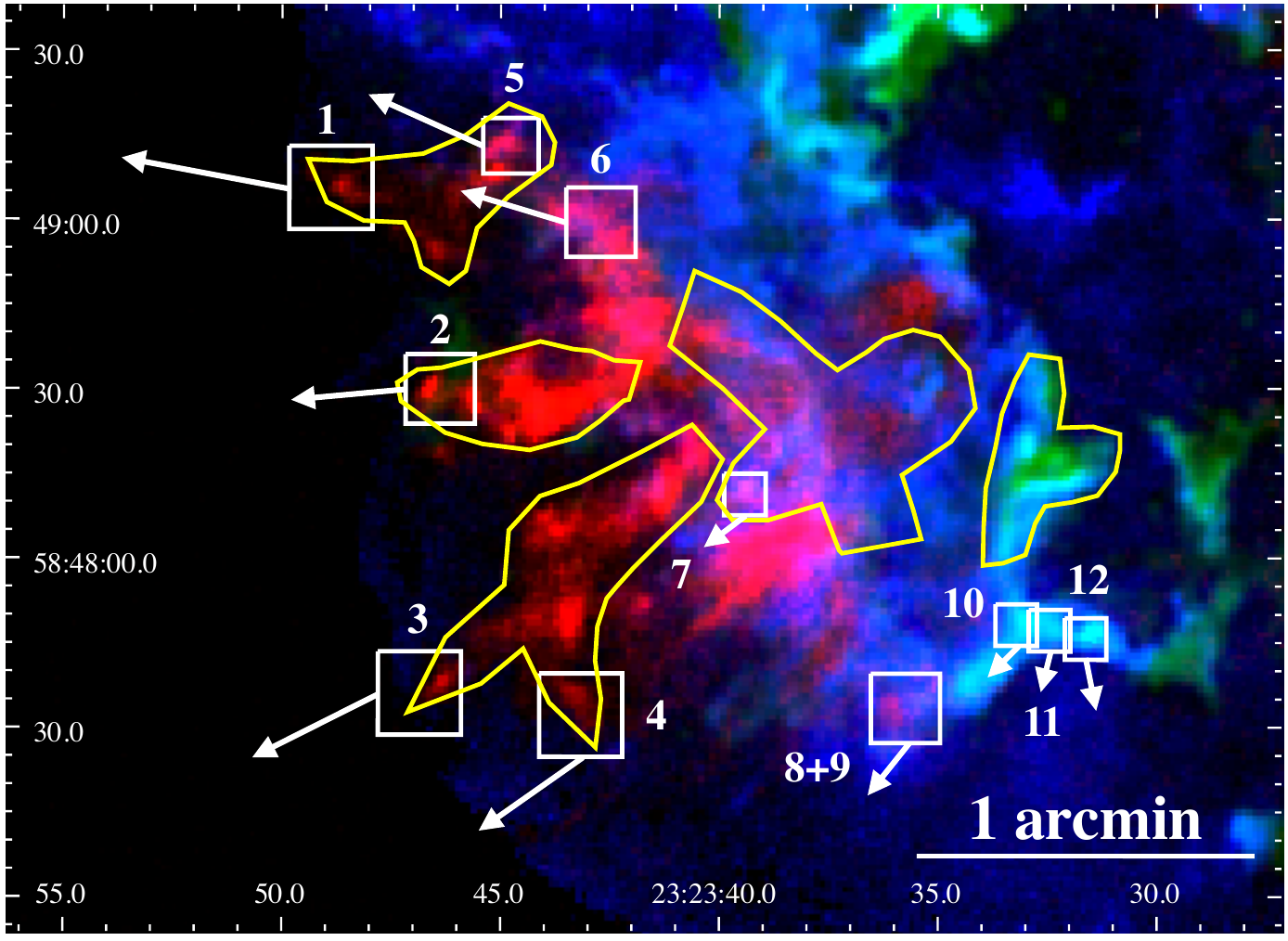}
 \caption{Enlarged image of the southeast area of Cas~A. The color scheme is the same as in Figure~1. 
 White boxes show the regions used for the proper motion analysis. White arrows indicate the direction and magnitude of the proper motion.}
\end{figure}

\subsection{Proper motion measurements}

We measure the proper motions of the Fe-rich and Si/O-rich ejecta in the southeastern region using some techniques to evaluate their kinematics in detail. Here we used X-ray images in the 0.5--1.7 keV band, which includes the Fe-L and O emissions, taken in 2000 and 2019.  

\begin{figure}[t] 
 \centering
 \includegraphics[width=8.5cm]{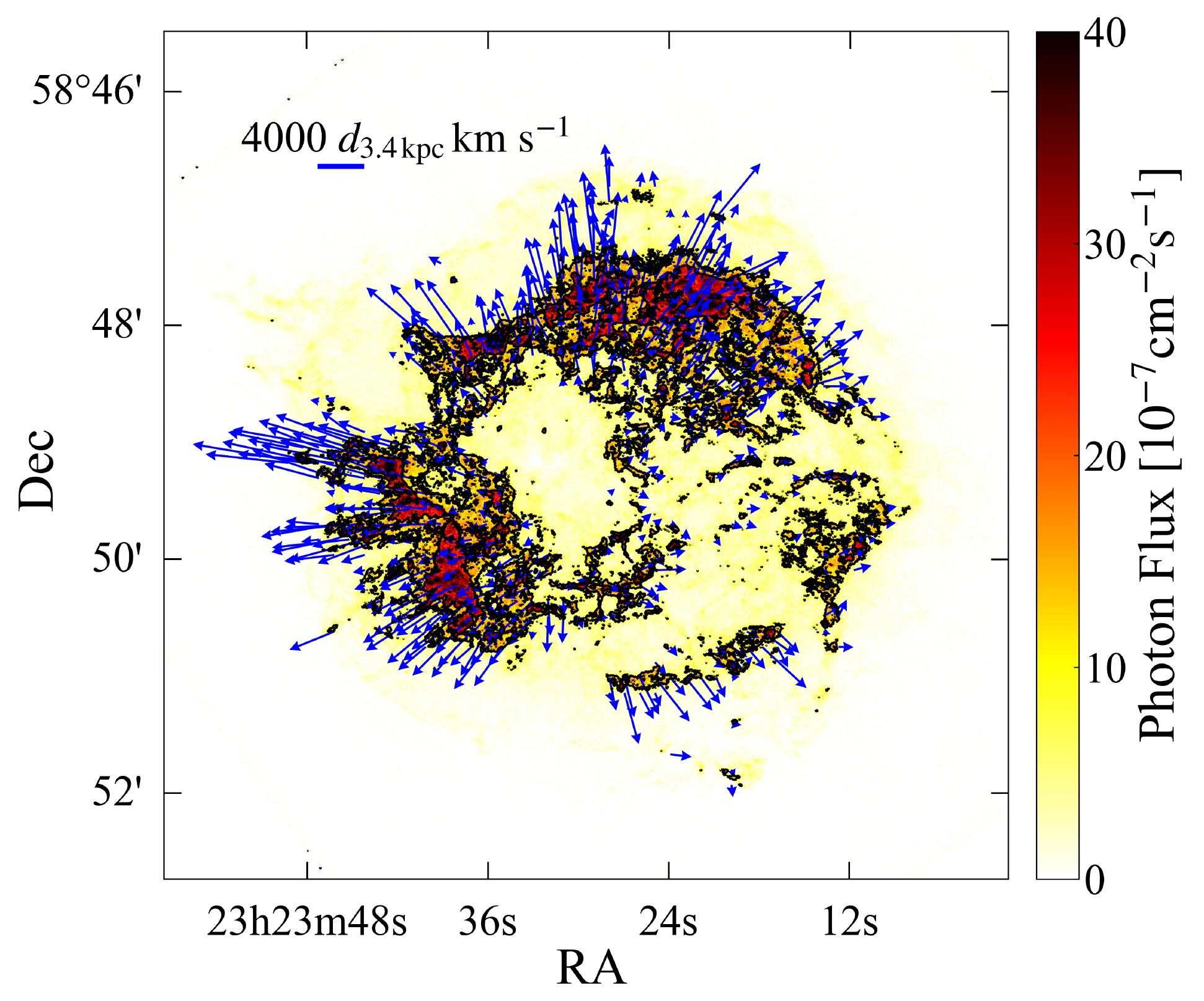}
 \caption{The X-ray image of Cas~A in 0.5-1.7 keV observed in 2000 overlaid with the velocity vectors obtained by optical flow. The scale bar indicates the 2D space velocity of 4,000 km s$^{-1}$ at the distance of 3.4 kpc.}
\end{figure}

To visualize expansions of local structures in Cas~A, we used a technique called ``optical flow'' \citep{farneback2003two}, which has been firstly applied to the remnant in \cite{2018ApJ...853...46S}. We used the {\tt calcOpticalFlowFarneback} function in OpenCV with the following arguments: {\tt pyr\_scale} = 0.5, which specifies the image scale to build pyramids for each image; {\tt levels} = 3; the number of pyramid layers; {\tt winsize} = 15; averaging window size; and {\tt poly\_n} = 5, which is the size of the pixel neighborhood used for the polynomial approximation. 
Assuming the distance of 3.4 kpc \citep{1995ApJ...440..706R}, the velocity vectors were calculated as shown in Figure~3. The moving structures at outer regions of the remnant have higher velocities, where the tip of the southeast region shows the velocities of $>$ 6,000 km s$^{-1}$.

We also measured the proper motions of 11 blob regions shown in Figure~2 using the maximum likelihood method \citep[e.g.,][]{2017ApJ...845..167S,2018ApJ...853...46S,2021ApJ...912..131T} to obtain the directions and magnitudes of the motions of each structure. 
Blob 5--12 in Figure~2 show the regions where the Doppler velocity measurements using High Energy Transmission Grating Spectrometer (HETGS) were performed in \cite{2013ApJ...769...64R}. The regions from Blob 1 to Blob 4 were defined as the outermost regions where Fe is abundant.
 As a result, the best-fit velocities in the plane of the sky (see 3rd and 4th rows in Table~2) were estimated to be $\sim$4,000--7,000 km s$^{-1}$ for the Fe-rich structures (Blob 1--6), $\sim$2,000--3,000 km s$^{-1}$ for the Fe/Si-rich structures (Blob 7--9), and $\sim$1,500--2,500 km s$^{-1}$ for the Si/O-rich structures (Blobs 10--12). The estimated velocities agree with those with the optical flow measurements. We found that the velocities of the Fe-rich structures in the plan of the sky are higher than those of the Si/O-rich structures. Even if we considered the line-of-sight (LoS) velocities obtained in \cite{2013ApJ...769...64R}, the velocities of the Fe-rich structures are higher than those of the Si/O-rich structures (see 5th and 6th rows in Table~2). On the other hand, the expansion index: $m$ ($r \propto t^{m}$, where $r$ and $t$ are the radius and age of the remnant, respectively) in the Si/O-rich structures estimated from the proper motion measurements are lower than that in the Fe-rich structures, which means that the Si/O-rich structures were subjected to stronger deceleration. In section 4.1, we discuss the deceleration in more detail.

\begin{deluxetable*}{cccccccc}[ht!]
\tablecaption{Velocity measurements in the regions shown in Figure~2 from Chandra data}
\tablehead{
 \colhead{Region\tablenotemark{a}} & \colhead{(R.A., Decl.)} & \twocolhead{Angular}  & \colhead{LoS} & \colhead{Total} &  \colhead{Expansion} & Free expansion \\[-1.8mm]
 \colhead{} & \colhead{} & \twocolhead{velocity}  & \colhead{velocity\tablenotemark{c}} & \colhead{} &  \colhead{index, $m$\tablenotemark{d}} & velocity\tablenotemark{d} \\
\colhead{}& \colhead{} & \colhead{(arcsec yr$^{-1}$)} & \colhead{(km s$^{-1}$)\tablenotemark{b}} & \colhead{(km s$^{-1}$)} &\colhead{(km s$^{-1}$) } &\colhead{}&\colhead{(km s$^{-1}$)}
}
\startdata
\hline
Blob 1 & \(23^{\mathrm{h}} 23^{\mathrm{m}} 48.^{\mathrm{s}} 89,58^{\circ} 49^{\prime} 05.^{\prime \prime} 59\) & $0.415 \pm 0.026 $ & $6700 \pm 410$ & - & - & 0.859 ± 0.053 & 7790 \\
Blob 2 & \(23^{\mathrm{h}} 23^{\mathrm{m}} 46.^{\mathrm{s}} 38,58^{\circ} 48^{\prime} 29.^{\prime \prime} 94\) & $0.282 \pm 0.026 $ & $4540 \pm 410$ & - & - & 0.666 ± 0.060 & 6820\\
Blob 3 & \(23^{\mathrm{h}} 23^{\mathrm{m}} 46.^{\mathrm{s}} 85,58^{\circ} 47^{\prime} 36.^{\prime \prime} 06\) & $0.342 \pm 0.026 $ & $5520 \pm 410$ & - & - & 0.722 ± 0.054 & 7640\\
Blob 4 & \(23^{\mathrm{h}} 23^{\mathrm{m}} 43.^{\mathrm{s}} 16,58^{\circ} 47^{\prime} 32.^{\prime \prime} 19\) & $0.311 \pm 0.026 $ & $5020 \pm 410$ & - & - & 0.769 ± 0.063 & 6520\\
Blob 5 & \(23^{\mathrm{h}} 23^{\mathrm{m}} 44.^{\mathrm{s}} 77,58^{\circ} 49^{\prime} 13.^{\prime \prime} 00\) & $0.308 \pm 0.026 $ & $4970 \pm 410$ & -1710 & 5260 & 0.781 ± 0.065 & 6370\\
Blob 6 & \(23^{\mathrm{h}} 23^{\mathrm{m}} 42.^{\mathrm{s}} 71,58^{\circ} 48^{\prime} 59.^{\prime \prime} 48\) & $0.266 \pm 0.026 $ & $4290 \pm 410$ & -1360 & 4500 & 0.781 ± 0.075 & 5500\\
Blob 7 & \(23^{\mathrm{h}} 23^{\mathrm{m}} 39.^{\mathrm{s}} 41,58^{\circ} 48^{\prime} 11.^{\prime \prime} 28\) & $0.128 \pm 0.026 $ & $2060 \pm 410$ & -540 & 2130 & 0.460 ± 0.092 & 4480\\
Blob 8 & \(23^{\mathrm{h}} 23^{\mathrm{m}} 35.^{\mathrm{s}} 74,58^{\circ} 47^{\prime} 33.^{\prime \prime} 40\) & $0.163 \pm 0.026 $ & $2630 \pm 410$ & -880 & 2780 & 0.607 ± 0.095 & 4340\\
Blob 9 & \(23^{\mathrm{h}} 23^{\mathrm{m}} 35.^{\mathrm{s}} 74,58^{\circ} 47^{\prime} 33.^{\prime \prime} 40\) & $0.163 \pm 0.026 $ & $2630 \pm 410$ & -810 & 2750 & 0.607 ± 0.095 & 4340\\
Blob 10 & \(23^{\mathrm{h}} 23^{\mathrm{m}} 33.^{\mathrm{s}} 21,58^{\circ} 47^{\prime} 48.^{\prime \prime} 17\) & $0.108 \pm 0.026 $ & $1750 \pm 410$ & -2340 & 2920 & 0.542 ± 0.128 & 3220\\
Blob 11 & \(23^{\mathrm{h}} 23^{\mathrm{m}} 32.^{\mathrm{s}} 45,58^{\circ} 47^{\prime} 47.^{\prime \prime} 18\) & $0.105 \pm 0.026 $ & $1700 \pm 410$ & -2300 & 2860 & 0.548 ± 0.133 & 3100\\
Blob 12 & \(23^{\mathrm{h}} 23^{\mathrm{m}} 31.^{\mathrm{s}} 62,58^{\circ} 47^{\prime} 45.^{\prime \prime} 71\) & $0.130 \pm 0.026 $ & $2100 \pm 410$ & -1360 & 2500 & 0.698 ± 0.137 & 3010\\
\hline
\enddata
\tablenotetext{}{All errors listed in the table represent statistical errors.}
\tablenotetext{a}{See Figure 2 for the regions. Each feature corresponds to the three regions in Figure 1: Blob 1-6 correspond to region A (Fe-rich), Blob 7-9 to region B (Fe/Si-rich), and Blob 10-12 to region C (Si/O-rich).}
\tablenotetext{b}{The distance to Cas~A is assumed to be 3.4 kpc \citep{1995ApJ...440..706R}.}
\tablenotetext{c}{The LoS velocities were measured with Chandra HETG in \cite{2013ApJ...769...64R}.}
\tablenotetext{d}{Only the values of Angular velocity were used.}
\label{tab:tab2}
\end{deluxetable*}

\begin{figure}[t] 
 \centering
 \includegraphics[width=8cm]{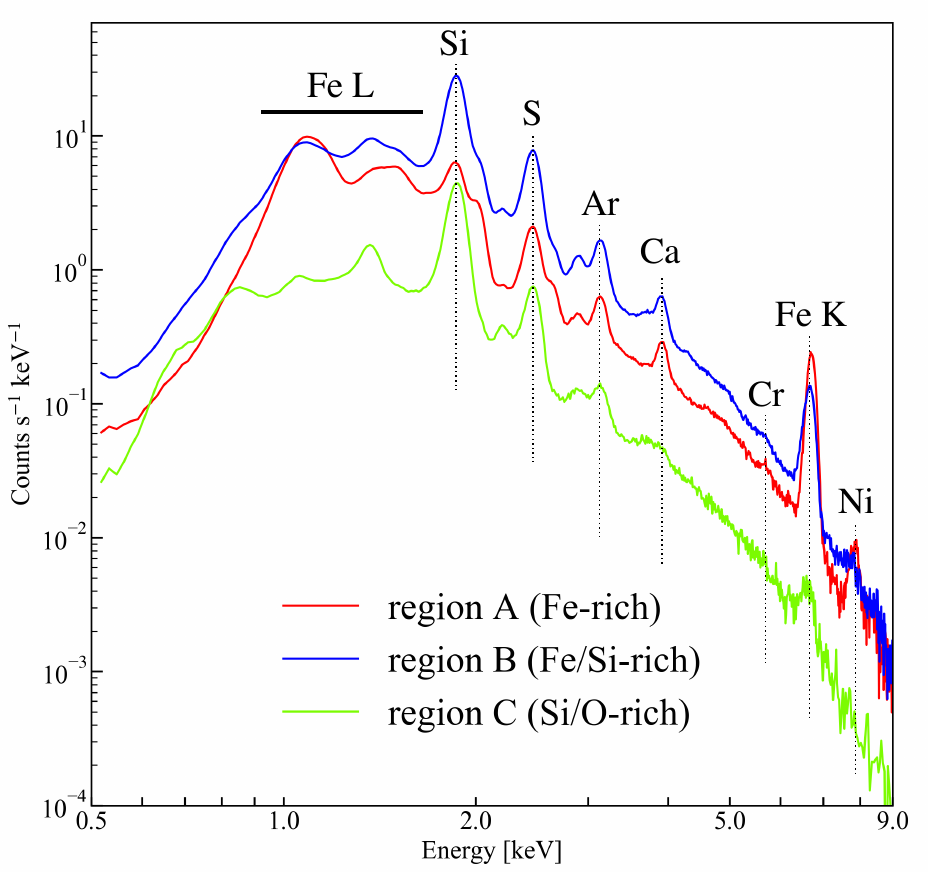}
 \caption{The Chandra spectra of Cas~A in 0.5--9.0 keV extracted from the regions defined in Figure~1 (red: region A, blue: region B, green: region C).}
\end{figure}

\begin{figure}[t] 
 \centering
 \includegraphics[width=8cm]{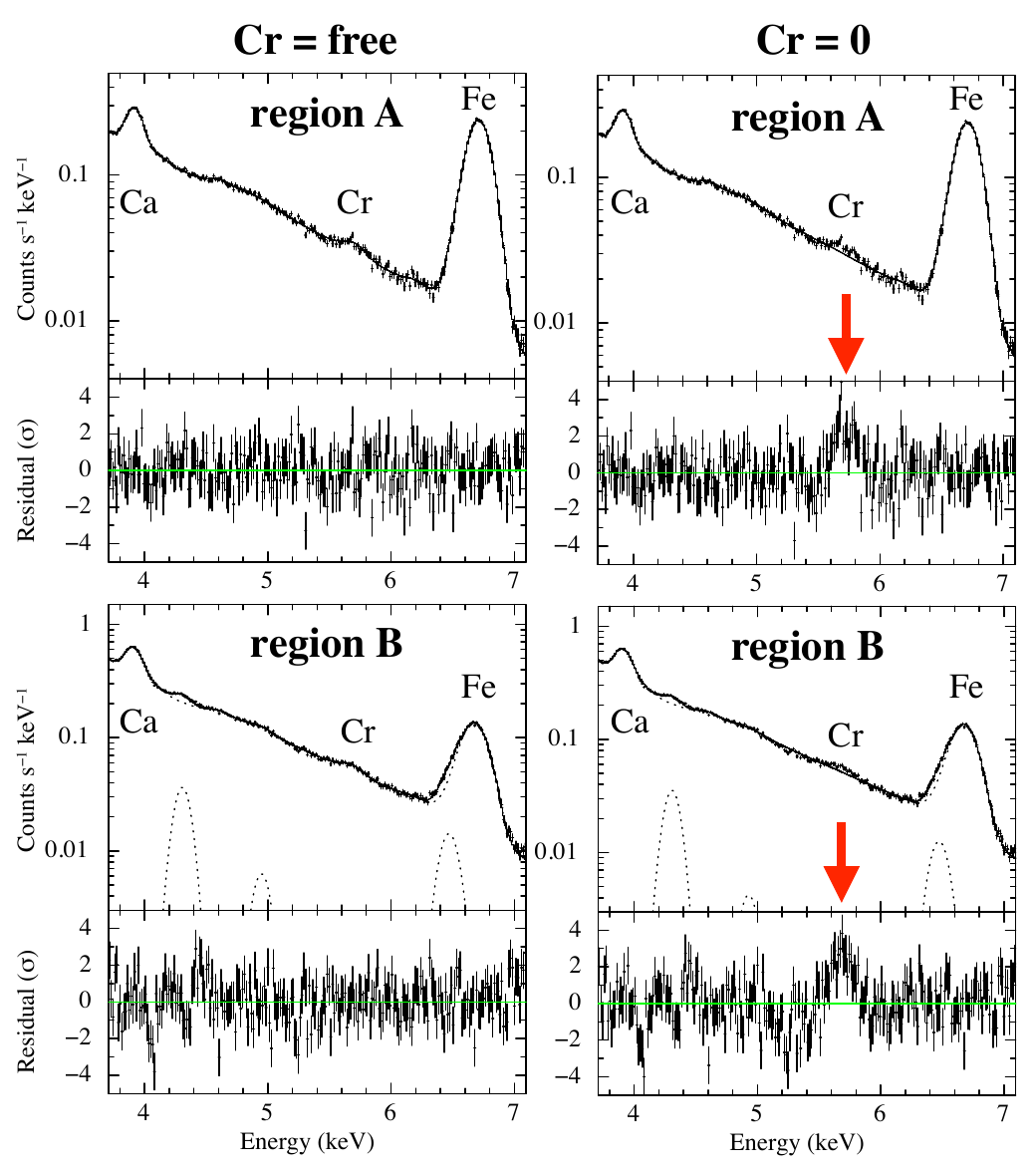}
 \caption{(Left row) X-ray spectra and best-fit models for region A (top) and B (bottom). (Right row) The same as the left row, but the abundances of Cr in the models are fixed at zero. At about 5.5--5.9 keV, the Cr-K$\alpha$ line emission can seen in large residuals.}
\end{figure}

\subsection{Spectral Analysis} 

In Figure~4, we showed X-ray spectra extracted from region A (Fe-rich region), B (Fe/Si-rich region), and C (Si/O-rich region). 
The comparison of the spectra shows that abundant elements are different from region to region, where elements that are prominent in the image appear as strong features in the spectra. We here focus on that the Si/Fe-rich structure (region B) is located just inside the Fe-rich structure (region A). In general, Si/Fe-rich ejecta (i.e., the production in the incomplete Si burning layer) should be produced at the outside of Fe-rich ejecta (i.e., the production in the complete Si burning layer) in the SN explosion. Therefore, the positional relationship between region A and region B may indicate an inversion, which has not been reported yet. 

To demonstrate the inversion between the complete and incomplete Si burning layers, we measured the abundances of Cr and Fe in region A and B by fitting the spectra in 3.7--7.1 keV. Since the magnitude of the Cr/Fe ratio changes between complete Si burning regime and incomplete Si burning regime \citep[e.g.,][]{2017ApJ...834..124Y,2020ApJ...890..104S,2021Natur.592..537S}, we chose Cr and Fe to quantify their mass ratio and identify the Si burning regime. The spectra were fitted with {\tt tbabs}, a model for interstellar absorption, and {\tt vvpshock}, a non-equilibrium ionized plasma model. We also used a {\tt gsmooth} model for expressing the thermal broadening and the Doppler effects. As a result, we found that the Cr/Fe mass ratio in region B is significantly higher than that in region A (Figure~5 \& Table~3), which implies that they were synthesized in different burning regimes (see section 4.2 for a detailed discussion). The best-fit parameters and the significance of the Cr detection are summarized in Table~3.

 In the spectral analysis, the hydrogen column density was fixed to $1.2 \times 10^{22}\, \mathrm{cm}^{-2}$, which is a typical value for the southeast region  (\cite{2003ApJ...597..362H,2021Natur.592..537S,2022arXiv220103753I}), while {\tt\(kT_{e}\)}, {\tt\(n_{e} t\)}, and normalization and abundances of Ca, Ti, Cr, Mn, and Fe are treated as free parameters and the other abundances of elements are frozen to the solar values in \cite{1989GeCoA..53..197A}. In region B, we found that the spectrum has significant contributions from pileup events due to the strong Si/S He$\alpha$ emissions (see the broken lines in the bottom panels in Figure~5). Therefore, we here added two Gaussian components ({\tt gauss} in XSPEC) at  $\sim$ 4.3 keV (Si He$\alpha$ + S He$\alpha$) and $\sim$ 4.9 keV (S He$\alpha$ $\times$ 2 or Si He$\alpha$ + Ar He$\alpha$) to express the line features from the pileup effect. The pileup effect would rarely affect our measurements of Cr and Fe abundances because line structures due to the pileup are expected to be too weak around these emission lines. Also, we added a Gaussian component at $\sim$6.5 keV to obtain a better fit for the Fe K$\alpha$ emission in region B. One possible reason for the failure to reproduce the Fe K$\alpha$ emission is that it is mixed with radiation from plasmas that are less ionized or moving away from the LoS direction than the majority of plasmas. The addition of this Gaussian model changes the estimated Cr/Fe ratio slightly (Table~3), but does not affect our interpretation in section 4.2.

\begin{deluxetable*}{cccccccc}[ht!]
\tablecaption{The best-fit parameters of spectral fittings for region A and B}
\tablehead{
 \colhead{Region} & \colhead{Model} & \colhead{Fe} &  \colhead{Cr} & \colhead{Mass Ratio [\%]} & \colhead{\(\chi^{2} \)/dof} & \colhead{\(\chi^{2} \)/dof} & Signficance\tablenotemark{a}\\
\colhead{} & \colhead{} & \colhead{} & \colhead{} & \colhead{(Cr/Fe)} & \colhead{Cr=0} & \colhead{Cr=free} & (Cr) }
\startdata
\hline
A & {\tt tbabs*gsmooth*vvpshock} & $7.18 ^{+0.29}_{-0.27}$ & $3.96 ^{+0.66}_{-0.66}$ & $0.51 ^{+0.11}_{-0.10}$ & 313.08/220 & 234.99/219 & 7.91 $\sigma$ \\
B & {\tt tbabs*gsmooth*(vvpshock + gauss$\times$2) }& $2.86 ^{+0.02}_{-0.05}$ & $3.82 ^{+0.51}_{-0.58}$ & $1.24 ^{+0.19}_{-0.20}$ & 496.79/214 & 385.42/213 & 7.35 $\sigma$\\
B & {\tt tbabs*gsmooth*(vvpshock + gauss$\times$3)} & $2.12 ^{+0.12}_{-0.16}$ &  $3.20 ^{+0.44}_{-0.43}$ & $1.40 ^{+0.32}_{-0.25}$ & 445.78/211 & 297.50/210 & 8.27 $\sigma$ \\
\hline
\enddata
\tablenotetext{a}{1$\sigma$ corresponds to a confidence interval of 68.3\%.}
\end{deluxetable*}

\section{Discussion} \label{sec:dis}

In previous sections, we investigated the kinematics and elemental compositions in the inverted ejecta layers in the southeastern region of Cas~A. Our results implied that the Fe-rich ejecta (i.e., the production by the complete Si burning regime) have higher velocities and are located on the outer side of the other materials that should be produced on the inner side of the SN explosion. These would be helpful to infer the inversion process during the explosion or remnant's evolution. In this section, we discuss these observational properties in more detail to understand the origin of the inverted layers. 

\begin{figure*}[t] 
 \centering
 \includegraphics[width=14.5cm]{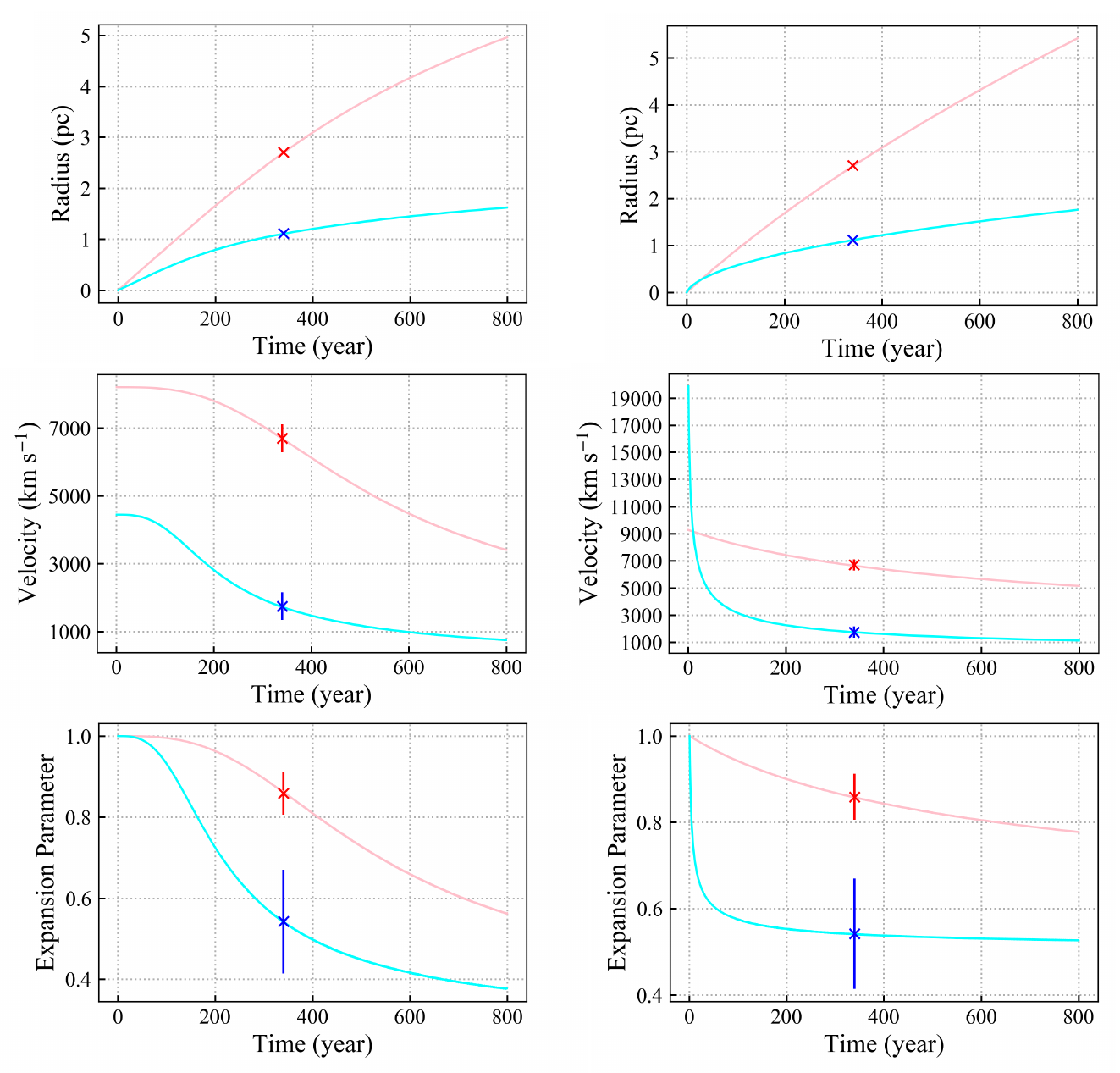}
 \caption{Time evolution of the radius (top panels), the velocity (middle panels), and the expansion index (bottom panels) from the momentum conservation law compared with the observed values for Blob 1 (red crosses) and Blob 10 (blue crosses). The density profiles of the surrounding medium are assumed to be uniform (left panels) and those formed by the wind of the progenitor star (right panels).
 Solid lines show our best-fit models for explaining the SNR parameters.}
 \label{fig6}
\end{figure*}

\subsection{The kinematics of the inverted layers}

In section 3.1, we found that the current velocities of the Fe-rich structures are higher than those of the Si/O-rich structures based on our proper motion measurements and the Doppler velocity measurements in \cite{2013ApJ...769...64R}. On the other hand, the estimations of the expansion index implied that the Si/O-rich structures seem to have undergone stronger deceleration than the Fe-rich structures. In the piston mechanism proposed in \cite{2010ApJ...725.2038D}, the Si/O-rich materials have higher velocities and larger radii at the beginning.
\cite{2021A&A...645A..66O} have demonstrated that the dense Fe-rich ejecta push out the less dense ejecta above (including Si), breaking through some of it \citep[see also][]{2016ApJ...822...22O}. Here, the low density Si/O-rich ejecta should experience strong deceleration, which may support our observations well. We here discuss the history of the expansion of each structure taking the deceleration into account.

We calculated the expansion index (see Table~2), which keeps the information of the ejecta deceleration, using the following equation:

\begin{equation}
m=\mu \times \frac{r}{t}
\end{equation}
where \(m\), \(\mu\), \(r\), and \(t\) indicate the expansion index, the observed proper motion, the distance from the explosion center, and the age of the remnant, respectively. Dividing the current velocity by the expansion index, we can obtain a free expansion velocity for traveling the current distance (shown in the 8th row of the Table~2), which would reflect the initial ejecta velocity roughly. As a result, we found that the free expansion velocities of the Fe-rich ejecta are still higher than those of the Si/O-rich ejecta.

In order to more accurately discuss the amount of deceleration, we solved for the time evolution of the supernova remnant assuming the momentum conservation law.
Here, we assume that the total mass of the remnant is increased by sweeping through the surrounding medium. We then considered two cases for the density profile of the surrounding medium: a constant density and a radial dependence on the density. In the former, as the simplest case, we assume a uniform surrounding medium with the density of $\rho_0$ as follows:
\begin{eqnarray}
\frac{v}{v_0}&=&\frac{1}{1+\frac{4 \pi \rho_{0} r_{1}^{3}}{3 M_{\mathrm{ej}}} \left(\frac{r}{r_{1}}\right)^{3}} 
 \equiv \frac{1}{1+A \left(\frac{r}{r_{1}}\right)^{3}}
\end{eqnarray}
where $M_{\mathrm{ej}}$, $v_0$, and $A = \frac{4 \pi \rho_{0} r_1^3}{3 M_{\mathrm{ej}}}$ are the ejecta mass, the initial ejecta velocity, and the mass ratio between the swept-up ISM and the ejecta as defined in Eq. (2) when the ejecta expands to $r_1$, respectively.
Here, $r_1$ is introduced to let the parameters be non-dimensional and it is arbitrarily set ($r_1$ = 1 pc is assumed throughout this paper). For example, if we set the typical parameters for Cas A as $M_{\mathrm{ej}} = 2-4 M_{\odot}$ \citep{2003ApJ...597..347L,2006ApJ...640..891Y} and $\rho_0 = 10^{-24}$ g cm$^{-3}$, we get $A = 3.09-1.55\ \times 10^{-2}$.

When solving the momentum conservation law, the initial parameters required for the calculation are only the initial velocity $v_0$ and the mass ratio $A$. We solved this equation to reproduce the observational properties obtained from the image analysis in the Fe-rich (Blob 1) and Si/O-rich (Blob 10) ejecta.

The left panels of Figure 6 show the time evolution of the radius, velocity, and expansion index obtained from our calculations. Here we set $v_0 = 8200$ km s$^{-1}$ and $A = 1.16\ \times 10^{-2}$ for the Fe-rich ejecta (red curve) and $v_0 = 4440$ km s$^{-1}$ and $A = 1.17$ for the Si/O-rich ejecta (blue curve), which reproduces the observational properties well. 
As a result, we found that the velocity of the Si/O-rich ejecta cannot be higher than that of the Fe-rich ejecta in the entire history of the expansion, even taking into account the statistical errors. Therefore, the inversion of the ejecta kinematics seems to have been produced at a very early stage of the remnant evolution or during the explosion, although there could be a large uncertainty in the initial velocity estimations using our simple calculations.

We would note that the mass ratio $A = 1.17$ for the Si/O-rich ejecta is almost two orders of magnitude larger than that for the Fe-rich ejecta, which implies that the Si/O-rich ejecta has encountered a denser ISM/CSM environment and/or had a lower ejecta density. Here the ejecta piston mechanism may explain this larger deceleration for the Si/O-rich ejecta \citep[e.g.,][]{2010ApJ...725.2038D,2021A&A...645A..66O}. In the simulations of \cite{2021A&A...645A..66O}, the density of Si ejecta at the outermost region in the early remnant stage have a much lower density than that of Fe ejecta (see top panels of Fig.~9 in the paper). The low-density Si ejecta undergoes a stronger slowdown and are eventually overtaken by the high-density Fe ejecta, which could be the main reason for the high mass ratio $A$ for the Si/O-rich ejecta in this region.

We then proceed to consider the latter model as a more realistic case, which reflects an SNR evolution in the progenitor’s wind bubble. The expansion parameters observed in Cas~A are known to be explained well with a pre-SN wind of a red supergiant, where the surrounding density profile is expressed with a function of $\rho \propto r^{-2}$ \citep[e.g.,][]{2003ApJ...593L..23C}. To calculate the SNR evolution in the progenitor's wind, we define the wind density profile of $\rho=A^{\prime} r^{-2}$.
For a steady wind, \(A^{\prime}\) can be written as \(A^{\prime}=\dot{M} / 4 \pi v_{w}\) using the mass loss rate \(\dot{M}\) and wind speed \(v_{w}\) \citep{2003ApJ...593L..23C}.
In this case, Eq. (2) can be rewritten as follows:

\begin{eqnarray}
\frac{v}{v_0}&=&\frac{1}{1+\frac{4 \pi A^{\prime} r_{1}}{3 M_{\mathrm{ej}}} \left(\frac{r}{r_{1}}\right)} 
 \equiv \frac{1}{1+B \left(\frac{r}{r_{1}}\right)}
\end{eqnarray}
where $B=\frac{4 \pi A^{\prime} r_{1}}{3 M_{\mathrm{ej}}}$, and the other parameters are the same as in Eq. (2).
The time evolution of the case with $v_0 = 9280$ km s$^{-1}$, $B = 0.148$ (red line) and with $v_0 = 21510$ km s$^{-1}$, $B = 10.2$ (blue line) is shown in the right panels of Figure 6, which reproduces well the observed results for Fe-rich and Si/O-rich ejecta, respectively.
Here, assuming $M_\mathrm{ej}=4 M_{\odot}$, the parameter $A^{\prime}$ of the equations that well represent the observed values for the Fe-rich and O-rich ejecta are $9.12 \times10^{13}$ g cm$^{-1}$ and $6.29 \times10^{15}$ g cm$^{-1}$, respectively.
The obtained values for the Fe-rich ejecta are consistent with $1.00 \times10^{14}$ g cm$^{-1}$, the value obtained in \citet{2003ApJ...593L..23C}.
This result suggests that the kinetic and spatial inversion should have occurred $\sim$10--30 yrs after the explosion even if assuming the expansion in the power-law density profile, which would be consistent with the piston mechanism proposed by \cite{2010ApJ...725.2038D,2021A&A...645A..66O}. Therefore, even assuming either of the two density profiles, our results support the inversion at the early stage of the remnant evolution.

\begin{figure}[t] 
 \centering
 \includegraphics[width=8.5cm]{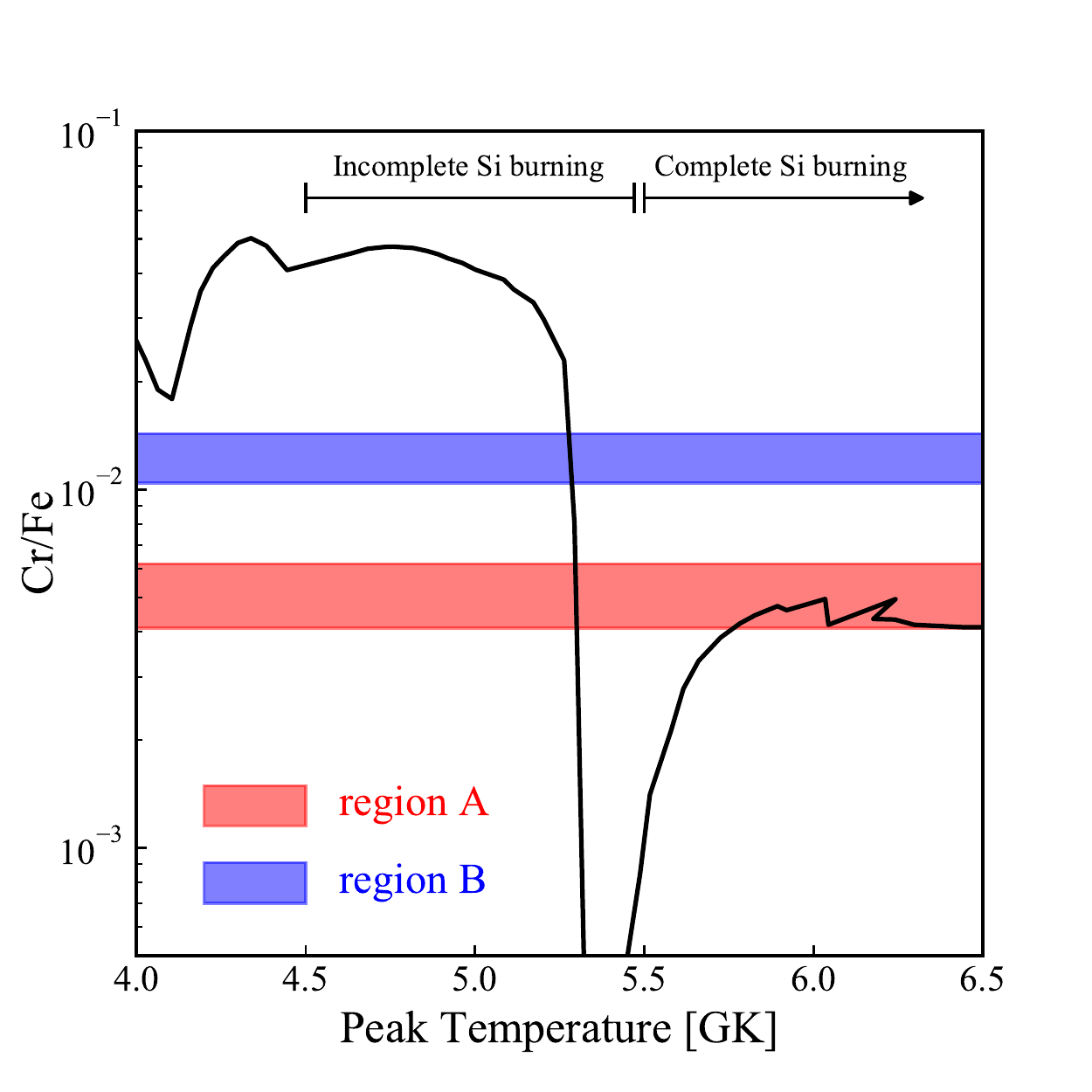}
 \caption{The relation between the peak temperature and the Cr/Fe mass ratio in a CC SN model. We assumed a 15 $M_\odot$ progenitor with sub-solar metallicity of $Z$ = 0.5 $Z_\odot$ and a high explosion energy of 3$\times$10$^{51}$ erg \citep[see][]{2020ApJ...893...49S,2021Natur.592..537S}. The red and blue areas indicate the observational values in region A and B, respectively.}
 \label{fig7}
\end{figure}

\subsection{Constraining the burning regimes}

In section 3.2, we found that the Cr/Fe mass ratio in the Fe/Si-rich region (region B) is larger than that in the Fe-rich region (region A), which implies that the complete Si burning layer is located just above the incomplete Si burning layer. In this section, we quantitatively discuss the observed mass ratios compared with a theoretical model to understand the nucleosynthetic origin of each region. 

The observed Cr/Fe mass ratio can be used for determining the peak temperature during the Si burning regime \citep{2017ApJ...834..124Y,2020ApJ...890..104S,2021Natur.592..537S,2022arXiv220103753I}.
Figure~7 shows the relation between the peak temperature during the nuclear burning and the Cr/Fe mass ratio in a 1D SN model used in \cite{2020ApJ...893...49S,2021Natur.592..537S}. Around $T_{\rm peak} =$ 5.5 GK, there is a transition between the incomplete and complete Si burning regimes, where the Cr/Fe mass ratio shows a difference of almost one order of magnitude. As for the Fe-rich region, the observed mass ratio (red area in the figure) falls within the complete Si burning regime as already reported in \cite{2021Natur.592..537S}. On the other hand, the Cr/Fe mass ratio in the Fe/Si-rich ejecta (blue area in the figure) is significantly out of the range of the complete Si burning regime, suggesting the incomplete Si burning origin ($T_{\rm peak} <$ 5.5 GK) for this structure.

The Cr/Fe mass ratio in the Fe/Si-rich ejecta is inconsistent with the production at the complete Si burning regime, though it is lower than the typical value at the incomplete Si burning around $T_{\rm peak} \approx$ 5 GK. One possibility is that most of the incomplete Si burning products are still not heated \citep[e.g.,][]{2020ApJ...904..115L}. Another is the strong $\alpha$-rich freezeout. \cite{2003ApJ...598.1163M} have shown that strong $\alpha$-rich freezeout realized by asymmetric explosions suppresses the production of elements produced by the incomplete Si burning regime, such as Cr and Mn.

To consider the amount of contamination from the ejecta produced by the complete Si burning would allow us to discuss the low Cr/Fe ratio in region B quantitatively. In the SN model we used, the typical Cr/Fe mass ratios by the complete and incomplete Si burning are about 0.004 and 0.04, respectively. Using these values, we have roughly estimated that the contribution from the complete Si burning ejecta in this region is $\sim$80\% to reproduce the observed ratio in region B. Interestingly, the contribution of the incomplete Si burning is small even in the inner Fe/Si-rich region, indicating most of Fe ($\gtrsim$ 80\%) in the entire east region are the complete Si burning (i.e., $\alpha$-rich freeze out) origin.

The inversion of these Si burning layers would be useful to discuss the formation process of the inverted layers in the southeastern region of Cas~A. Currently, the ejecta pistons during the SNR evolution would be the reasonable mechanism to explain it. On the other hand, the inversion of the adjacent Si burning layers should occur before the inversion between Fe and Si ejecta, which may indicate the ejecta overturning during the explosion. In addition, such strong asymmetry could suppress the incomplete Si burning products \citep{2003ApJ...598.1163M}, and this picture supports our results well. Further theoretical studies comparing with our observations will be helpful to understand the origin of the layered structure and the strong asymmetry during the SN explosion.

\section{Summary} \label{sec:sum}

Central strong activities are widely believed to be an important process in the explosion of massive stars. X-ray observations of the Cassiopeia~A supernova remnant have indicated a possibility of ``overturning'' of the ejecta during the explosion, which is the best target to test such a strong asymmetry.
In this paper, we have investigated the kinematic and nucleosynthetic properties of the inverted ejecta layers to understand its formation process and the SN central activity using the Chandra observations.

 Based on imaging and spectral analysis, we found that the 3D velocities of Fe- and Si/O-rich ejecta are $>$4,500 km s$^{-1}$ and $\sim$2,000--3,000 km s$^{-1}$, respectively. Even considering the deceleration, the velocity of the Si/O-rich ejecta is rarely higher than that of the Fe-rich ejecta in the entire history of the expansion.

In addition, we measured the Cr/Fe mass ratios in the ejecta layers to constrain their burning regime. We found that the mass ratio in the Fe/Si-rich region located at the inside of the Fe-rich region is significantly higher than that in the Fe-rich region. This means that the complete Si burning layer is located just outside the incomplete Si burning layer. All the results suggest that the ejecta overturning has been produced at the early stages of the remnant's evolution or during the supernova explosion of Cassiopeia~A. In the future, combined with theoretical modelings, our results will lead to an understanding of the origin of this spatial inversion of the ejecta layers.

\section*{acknowledgements}
We thank Dr. Takashi Yoshida and Prof. Hideyuki Umeda for providing the nucleosynthesis data. This work was supported by JSPS KAKENHI Grant Numbers 18H03722, 19K14749, 20H01941, and 20K20527.





\end{document}